\documentclass{cernrep} 
\usepackage{texnames}
\usepackage[T1]{fontenc}
\usepackage[bookmarks, colorlinks=true, linktoc=page, pdftex, linkcolor=red, citecolor=red, urlcolor=red]{hyperref}
\usepackage[usenames,dvipsnames,svgnames]{xcolor}

\renewcommand{\Pb}{{\rm Pb}}
\newcommand{\pp}{{{\rm p}-{\rm p}}}

\begin{document}
\title{
LHC limits on axion-like particles from heavy-ion collisions 
}
 
\author{Simon Knapen$^{1,2}$, Tongyan Lin$^{1,2,3}$, Hou Keong Lou$^{1,2}$ and Tom Melia$^{1,2,4}$}
\institute{$^{1}$ Department of Physics, University of California, Berkeley, California 94720, USA \\
$^{2}$ Theoretical Physics Group, Lawrence Berkeley National Laboratory, Berkeley, California 94720, USA\\
$^{3}$ Department of Physics, University of California, San Diego, California 92093, USA \\
$^{4}$ Kavli Institute for the Physics and Mathematics of the Universe (WPI), University of Tokyo Institutes for Advanced Study, University of Tokyo, Kashiwa 277-8583, Japan}

\begin{abstract}
In these proceedings we use recent LHC heavy-ion data to set a limit on axion-like particles coupling to electromagnetism with mass in the range 10-100 GeV. We recast ATLAS data as per the strategy proposed in \cite{Knapen:2016moh}, and find results in-line with the projections given there.
\end{abstract}


\maketitle 

\section{Introduction}

The LHC has completed its highest luminosity heavy-ion collision run (Pb-Pb),  with ATLAS, CMS and ALICE all recording data at a centre-of-mass energy per nucleon of $\sqrt{s_{NN}}=5.02$ TeV. In previous work \cite{Knapen:2016moh} we showed that the large charge of the lead ions ($Z=82$) results in a huge $Z^4$ coherent enhancement in the exclusive production of axion-like particles (ALPs) that couple to electromagnetism, which can lead to competitive limits for ALPs. This proceeding is an update to our previous work; we recast the analysis of the ATLAS $480\, \mu b^{-1}$ data set \cite{Aaboud:2017bwk} to provide limits on ALPs in the mass region $10\,\text{GeV}< m_a<100\,\text{GeV}$. In line with the projections in \cite{Knapen:2016moh}, we find that the LHC heavy-ion data provides the strongest limits to date in this parameter range. While the physics potential of exclusive heavy ion collisions has been known for decades \cite{Baur1988,Drees:1989vq,Papageorgiu:1988yg}, to our knowledge this represents the first time LHC heavy-ion data sets the most stringent limit on a specific beyond the Standard Model physics scenario.

Ultra-peripheral collisions (UPCs) are quasi-elastic processes where the impact parameter is much greater than the ion radius (see {\it e.g.} Refs.~\cite{Baur:2001jj, Bertulani:2005ru, Baltz:2007kq}). The ions remain (largely) intact, and there is a large rapidity gap between any produced particle and the beam-line with very little detector activity. This clean environment, along with the $Z^4$ enhanced signal rate, provides a low background ALP search channel that can perform better than searches using the \pp{} run.

\begin{figure}[t!]
\centering
\includegraphics[width=6cm]{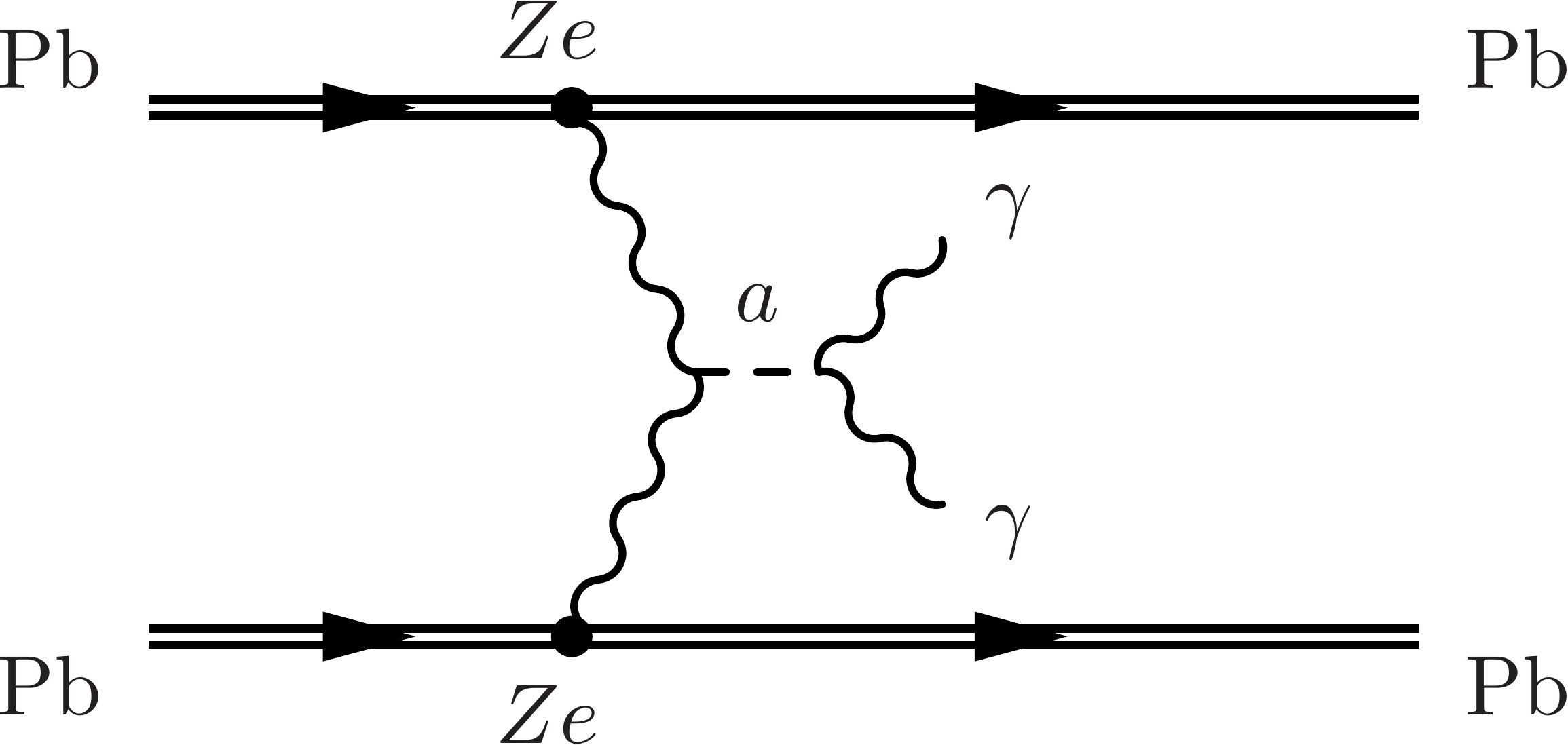}
\caption{Exclusive ALP production in ultra-peripheral Pb-Pb collisions.}
\label{fig:fey_alp2}
\end{figure}

The production of an ALP in a UPC proceeds via photon fusion---see Fig.~\ref{fig:fey_alp2}---where we consider a Lagrangian of the form
\begin{equation}\label{eq:lag2}
  \mathcal{L}_a = 
  \frac{1}{2}(\partial a)^2 -\frac{1}{2}m_a^2 a^2- \frac{1}{4}\frac{a}{\Lambda} F \widetilde{F}\,,
\end{equation}
where $\widetilde{F}^{\mu\nu}\equiv\epsilon^{\mu\nu\rho\sigma}F_{\rho\sigma}/2$, and 
with $a$ being the pseudoscalar ALP of mass $m_a$ which couples to electromagnetism via the dimensionful coupling $1/\Lambda$. Such a coupling can be obtained through the $SU(2)_L$ invariant operator $-a B \widetilde{B}/(4 \cos^2 \theta_W \Lambda)$ where $B$ is the hypercharge field strength. Polarization effects of the incoming photons can lead to different scalar and pseudoscalar production rates, but the effects are relatively small when integrating over all impact parameters~\cite{Greiner:1990aq,Baur:1990fx}. Our limits therefore apply for scalar particles through the replacement $\tilde F$($\tilde B$) $\to$ $F$($B$) in Eq.~\eqref{eq:lag2}.

\begin{figure}[th!]
  \centering
  \includegraphics[width=0.47\textwidth]{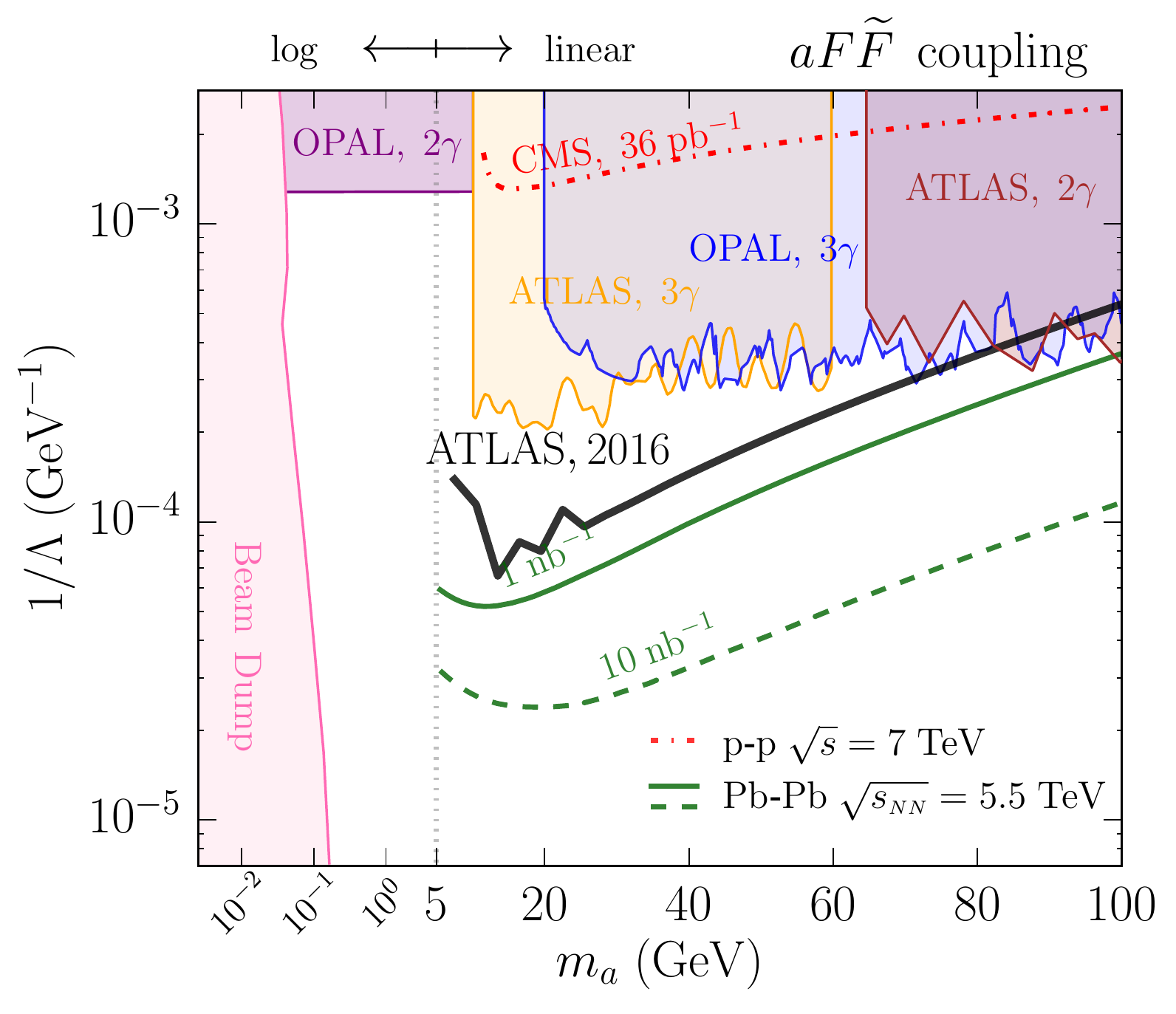}
  \includegraphics[width=0.47\textwidth]{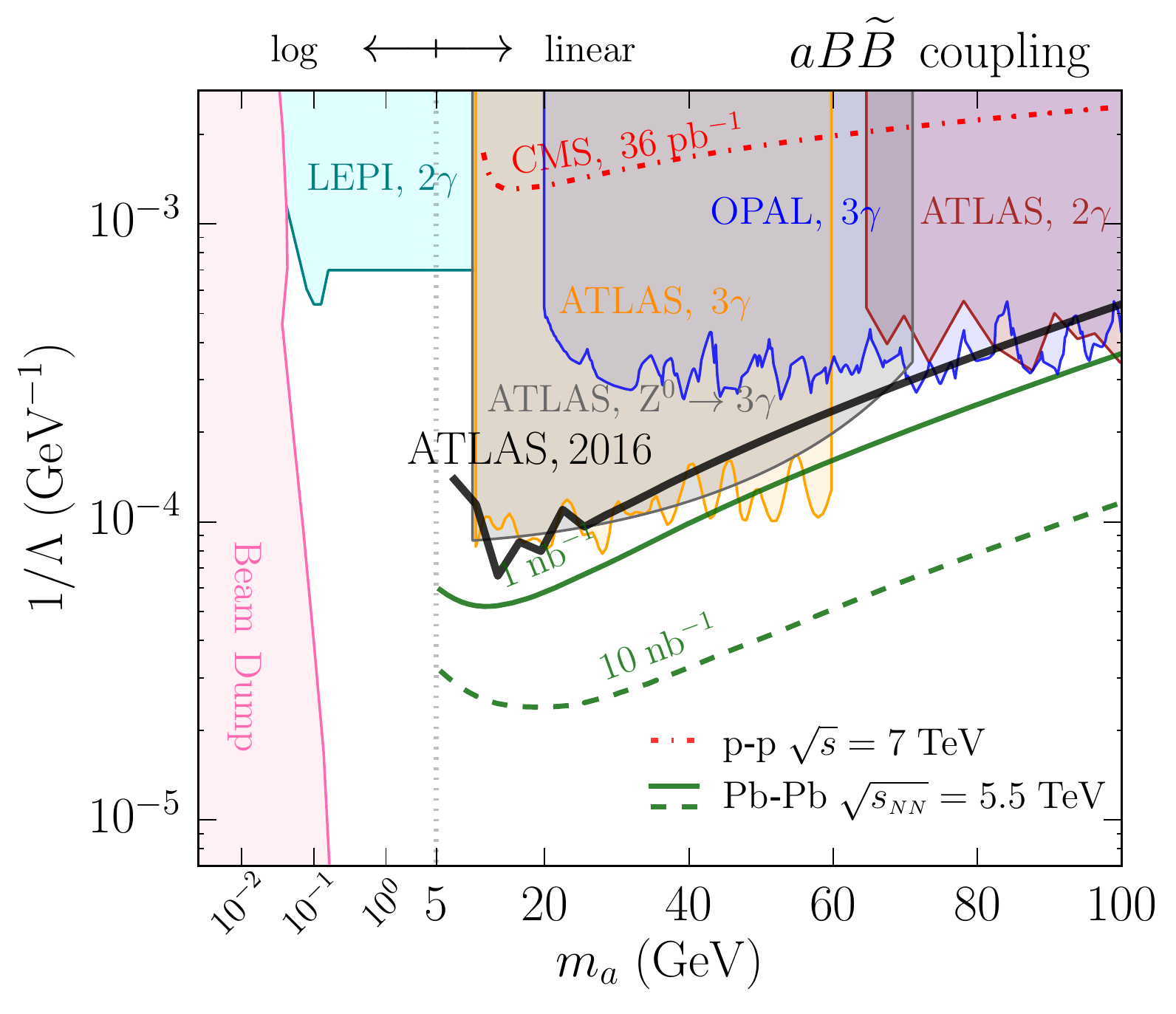}
  \caption{\label{fig:aFF} \emph{Left:} We show 95$\%$ exclusion limits on the operator $\frac{1}{4}\frac{1}{\Lambda}a F\tilde F$ using recent ATLAS results on heavy-ion UPCs~\cite{Aaboud:2017bwk} (solid black line). The expected sensitivity assuming a luminosity of 1\,nb$^{-1}$ (10\,nb$^{-1}$) is shown in solid (dashed) green.  For comparison, we also give the analogous limit from 36\,pb$^{-1}$ of exclusive p-p collisions \cite{Chatrchyan:2012tv} (red dot-dash). Remaining exclusion limits are recast from LEP II (OPAL $2 \gamma$,  $3 \gamma$) \cite{Abbiendi:2002je} and from the LHC (ATLAS 2$\gamma$, 3$\gamma$) \cite{Aad:2014ioa,Aad:2015bua} (see \cite{Knapen:2016moh} for details). 
\emph{Right}: The corresponding results for the operator $\frac{1}{4\cos^2 \theta_W}\frac{1}{\Lambda}a B\tilde B$.   The LEP I, $2\gamma$ (teal shaded) limit was obtained from \cite{Jaeckel:2015jla}.
 }
\end{figure}

Here we recap some of the details of our treatment in \cite{Knapen:2016moh}. The ALP parameter space is already substantially constrained by cosmological and astrophysical observations, as well as by a broad range of collider and intensity frontier experiments (see {\it e.g.}~\cite{Essig:2013lka,Bauer:2017ris} for reviews and recent results). In the regime of interest for UPCs (${1 \; {\rm GeV} \lesssim m_a\lesssim 100 \; {\rm GeV}}$), the existing constraints however come from LEP and LHC \cite{Mimasu:2014nea, Jaeckel:2015jla,Jaeckel:2012yz}. In Fig.~\ref{fig:aFF}, we show the expected sensitivity from performing a bump hunt in $m_{\gamma\gamma}$ for UPCs, assuming a luminosity for the current (1\,nb$^{-1}$) and the high luminosity (10\,nb$^{-1}$) Pb-Pb runs.\footnote{Limits from the p-Pb runs are not competitive despite their higher luminosity, because of the less advantageous $Z^2$ scaling of the production rate.  Collisions with lighter elements, {\it e.g.~}Ar-Ar, may set relevant limits if the luminosity could be enhanced by two to three orders of magnitude, as compared to current Pb-Pb run.} For each mass point we computed the expected Poisson limit~\cite{Agashe:2014kda}. The dominant backgrounds are estimated to be light-by-light scattering \cite{Baur1988} and fake photons from electrons, and become negligible for $m_{\gamma\gamma} \gtrsim 20$ GeV. In the region which there is background, we assume the entire signal falls into a bin of width 1 GeV. The signal selection criteria in this case are $E_T > 2$ GeV and $|\eta| < 2.5$ for the two photons and $|\phi_{\gamma\gamma} - \pi| < 0.04$. The analogous limit from the exclusive p-p analysis performed by CMS~\cite{Chatrchyan:2012tv} is also shown, which is very weak due to low photon luminosities. For the $F\tilde F$ operator the heavy-ion limits are significantly stronger, whereas for the $B\tilde B$ operator, traditional \pp{} collider limits are enhanced due to additional production channels through the $Z$ coupling.

%

Light-by-light scattering has been measured by the ATLAS collaboration~\cite{Aaboud:2017bwk}, and the results were consistent with our estimates and those in earlier computations \cite{d'Enterria:2013yra, Klusek-Gawenda:2016euz,Klusek-Gawenda:2016nuo}. Using the observed $m_{\gamma \gamma}$ spectrum, we then derive an observed limit on ALPs for $F\widetilde F$ and $B\widetilde B$ couplings, which are shown in black in Fig.~\ref{fig:aFF}. In detail, we generated Monte Carlo samples for the ALP signal using a modified version of the {\tt STARlight} code~\cite{starlight},\footnote{Our patch for ALP production is now included in the latest {\tt STARlight} release.} which assigns a small virtuality to the photons and as such leads to a typical $p^{\gamma\gamma}_{T}\lesssim 100\,$MeV for the recoil of the $\gamma\gamma$-system. We then follow the ATLAS analysis and apply the following selection cuts on the signal:
\begin{enumerate}
\item Require exactly two photons with $E_T > 3$ GeV and $|\eta| < 2.4$ 
\item Demand $|\phi_{\gamma \gamma} -\pi|< 0.03$, where $\phi_{\gamma \gamma}$ is the azimuthal angle between the two photons
\end{enumerate}
The signal efficiency is $\sim$70\% near threshold and becomes fully efficient if the sum of the photon energies exceeds 9 GeV. The selection criteria are slightly different from our previous theoretical analysis, however we note that only the larger $E_T$ cut leads to noticeable changes for the efficiencies. Given that we do not model photon identification at the detector level, we apply an extra total reconstruction efficiency of 90\%, which roughly takes into account the per-photon ID efficiency of 95\% measured by ATLAS. 

The $m_{\gamma \gamma}$ spectrum measured by ATLAS is plotted in bin-widths of $3 $ GeV, starting at $m_{\gamma \gamma} = 6$ GeV. For our exclusion, we generated samples with $m_{\gamma \gamma} = 7, 10, 13, 16, ... $ GeV, and assume that all the events are contained in their respective bins after final selection. We further assume that ATLAS did not observe any events with $m_{\gamma \gamma} \gtrsim 30 $ GeV. The 95\% exclusion limits on the coupling $1/\Lambda$ are obtained assuming only statistical uncertainties. A more detailed CL$_s$ analysis that includes a proper treatment of systematics would yield slightly more conservative limits, and we encourage the experimental community to include such an analysis as it is beyond the scope of our simulation framework.

In summary, we have found that heavy-ion collisions at the LHC can provide the best limits on ALP-photon couplings for $7\,\mathrm{GeV}<m_a<100\,\mathrm{GeV}$, confirming our previous estimates. The very large photon flux and extremely clean event environment in heavy-ion UPCs provides a rather unique opportunity to search for BSM physics.

\section*{Acknowledgements}

We thank Bob Cahn, Lucian Harland-Lang, Yonit Hochberg, Joerg Jaeckel, Spencer Klein, Hitoshi Murayama, Michele Papucci, Dean Robinson, Sevil Salur, and Daniel Tapia Takaki for useful conversations. 
SK and TL also thank the participants of the 3rd NPKI workshop in Seoul for useful comments and discussions. 
This work was performed under DOE contract DE-AC02-05CH11231 and NSF grant PHY-1316783.

\bibliographystyle{utphys}
\bibliography{alps}
  
\end{document}